\documentclass[
reprint,
amsmath,amssymb,aps,
]{revtex4-1}

\newtheorem{theorem}{Theorem}
\newtheorem{proposition}{Proposition}
\newtheorem{lemma}{Lemma}
\newtheorem{corollary}{Corollary}
\newtheorem{remark}{Remark}

\def\ba{\begin{array}}
\def\ea{\end{array}}
\def\be{\begin{equation}}
\def\ee{\end{equation}}

\def\ds{\displaystyle}

\def\i{{\bf i}}

\def\v{{\bf v}}

\def\0{{\bf 0}}
\def\1{{\bf 1}}
\def\2{{\bf 2}}
\def\3{{\bf 3}}
\def\4{{\bf 4}}
\def\5{{\bf 5}}
\def\6{{\bf 6}}
\def\7{{\bf 7}}
\def\8{{\bf 8}}
\def\9{{\bf 9}}

\usepackage{graphicx}\usepackage{dcolumn}\usepackage{bm}\usepackage{amsmath}\usepackage{amscd}\usepackage{amsfonts}

\def\bt{\begin{theorem}}
\def\et{\end{theorem}}
\def\bp{\begin{proposition}}
\def\ep{\end{proposition}}
\def\bc{\begin{corollary}}
\def\ec{\end{corollary}}
\def\bo{\begin{proof}}
\def\eo{\end{proof}}
\def\bx{\begin{example}}
\def\ex{\end{example}}
\def\br{\begin{remark}}
\def\er{\end{remark}}
\def\bl{\begin{lemma}}
\def\el{\end{lemma}}

\begin{document}

\preprint{APS/123-QED}
\title{Quantum Algorithm to Cubic Spline Interpolation}
\author{Changpeng Shao}
\email{cpshao@amss.ac.cn}
\affiliation{Academy of Mathematics and Systems Science, Chinese Academy of Sciences\\ Beijing 100190, China}
\date{\today}
\begin{abstract}
HHL algorithm \cite{harrow} to solve linear system is a powerful and efficient quantum technique to deal with many matrix operations (such as matrix multiplication, powers and inversion). It inspires many applications in quantum machine learning \cite{biamonte, dunjko}. However, due to the restrictions of HHL algorithm itself, many quantum machine learning algorithms also share one or two restrictions. The most common restrictions include quantum state preparation, condition number and Hamiltonian simulation. In this work,
we first give an efficient quantum algorithm to achieve quantum state preparation, which actually achieves an exponential speedup than the algorithms given in \cite{clader,lloyd13}.
Then we provide an application of HHL algorithm in cubic spline interpolation problem. We will show that in this problem, the condition number is small, the preparation of quantum state is efficient based on the new algorithm we proposed and the Hamiltonian simulation is efficiently implemented. So the quantum algorithm obtained by HHL algorithm towards this problem actually achieves an exponential speedup than any classical algorithm with no restrictions.
This can be viewed as another application of HHL algorithm with no restrictions after the work of Clader et al \cite{clader} in studying electromagnetic scattering cross-section.
\end{abstract}

\pacs{Valid PACS appear here}
\maketitle

\section{Introduction}

HHL algorithm \cite{harrow}  to solve linear system $Ax=b$ is an important quantum linear algebra based subroutine of many quantum algorithms to machine learning problems, such as quantum principal analysis \cite{lloyd13}, support vector machine \cite{rebentros14}, neural network \cite{rebentros17-Network}, data fitting \cite{schuld, wang, wiebe}, optimization \cite{rebentros17b}, Boolean equations solving \cite{chen}, to name a few.
However, because of the restrictions of HHL about quantum state preparation of $b$, Hamiltonian simulation $e^{-\i \widetilde{A}t}$ and the dependence on condition number of $A$, where $\widetilde{A}=\left[
   \begin{array}{cc}
     0 & A \\
     A^\dag & 0 \\
   \end{array}
 \right]$, these quantum machine learning algorithms also possess several of the restrictions.
Quantum state of $b$ can be prepared efficiently when $b$ is sparse or relatively uniform, i.e., without a few entries that are vastly larger than the others \cite{clader}.
And Hamiltonian simulation is efficient when $A$ is sparse \cite{berry, berry1} or low rank \cite{rebentros16}.
However, the condition number of $A$ is not easy to estimate generally. These restrictions will be the main concerns in studying cubic spline interpolation problem in this work.

On one hand, finding more applications of HHL algorithm is an important task that can provide us more examples that quantum computer can speedup. Since under certain conditions, such quantum algorithms will achieve exponential speedup than all the classical algorithms. On the other hand, finding more applications of HHL algorithm with fewer or no restrictions are convincing to show the potential power of quantum computer. To the best of my knowledge, one such application of HHL algorithm seems to be obtained by Clader et al \cite{clader} at 2013 in studying electromagnetic scattering cross-section problem via finite element method.

A typical application of HHL algorithm is linear regression (i.e., data fitting) \cite{schuld, wang, wiebe}, since HHL algorithm actually obtains the least square solution. Although, HHL algorithm only obtain the quantum state of the solution, it is enough to do the prediction on new data by swap test \cite{buhrman}. Generally, all the three restrictions discussed above are unavoidable in data fitting. However, from the viewpoint of practicality, locally weighted linear regression is more useful. It is simple and effective than polynomial regression when linear regression is not enough. Also when considering about Hamiltonian simulation, locally weighted linear regression is more suitable to study by quantum computer \cite{kerenidis}, since it corresponds to a low rank linear system, whose Hamiltonian simulation can be implemented efficiently \cite{rebentros16, lloyd13, wossnig}. So, in locally weighted linear regression, the only restrictions we may encounter are quantum state preparation and the condition number.

A closely related research topic is polynomial interpolation or approximation \cite{burden}. Global interpolation method like Lagrange interpolation, Newton interpolation or Hermite interpolation often generates a polynomial with high degree and contains an expensive cost in calculation. Sometimes they are even unstable and inaccurate. Local interpolation method includes piecewise linear interpolation, cubic Hermite interpolation, cubic spline interpolation and so on. Among which cubic spline interpolation performs pretty well than others. Cubic spline interpolation is smoother than cubic Hermite interpolation, also it can avoid Runge's phenomenon. It is a stable interpolation method, which contains a high rate of convergence and a low cost of computation. Also it is very useful both in practice, such as in signal processing, image processing, curve fitting, chemical physics and so on.

More importantly (from the point of quantum computer), it reduces to solve certain linear systems, whose coefficient matrices are diagonally dominant and tridiagonal. So a direct result of \cite{berry, berry1}  is that the Hamiltonian simulation relates to the coefficient matrices of these linear systems is efficient.
Furthermore, we will show that the condition numbers of the coefficient matrices are bounded by a small constant (i.e., $4\sqrt{2}$).

As for the quantum state preparation problem, in this work, we will propose a new efficient quantum algorithm solve it, which achieves an exponential speedup than the algorithms of \cite{clader,lloyd13}. Although, it can not solve the quantum state preparation problem efficiently for all cases, the exponential speedup provides us more evidences to trust that the quantum state preparation problem can be solved efficiently in many practical problems, such as the locally weighted linear regression or the cubic spline interpolation considered in this work.
Based on this new efficient quantum algorithm, we will show that the required quantum states can be prepared efficiently.
All these results imply that HHL algorithm can play a positive role in this problem with no restrictions.
Moreover, just like data fitting problem, when obtaining the quantum state of the solution by HHL algorithm, the evaluation on new data can be resolved easily by swap test. Also, the quantum state of the new data only contains two nonzero entries, which can be prepared efficiently. Therefore, cubic spline interpolation seems to be a very ``clean" application of HHL algorithm.

The structure of this work is as follows:
Section \ref{HHL Algorithm and Swap Test} mainly focus on the introduction of certain necessary techniques of quantum computer that will be used in this work.
First, we will give a comprehensive analysis about HHL algorithm, which can also be regarded  as a short review of HHL algorithm.
Then we briefly introduce swap test.
In section \ref{Quantum State Preparation}, we will give an efficient method to solve the quantum state preparation problem.
Section \ref{Preliminaries of Cubic Spline Interpolation} devotes to present some preliminaries about cubic spline interpolation.
In section \ref{Bounds on Condition Number}, an analysis about the upper bound of the condition numbers about the linear systems appeared in the cubic spline interpolation will be given.
Finally, in section \ref{Quantum Cubic Spline Interpolation}, we apply HHL algorithm to solve the cubic spline interpolation problem with exponential speedup.

{\bf Notations.} For any matrix $A=(a_{ij})_{n\times n}$, its Frobenius norm is defined as $\|A\|_F=\sqrt{\sum_{i,j}|a_{ij}|^2}$. In this paper $\|\cdot\|$ always refers to the 2-norm of vectors and $\i$ refers to imaginary unit $\sqrt{-1}$ of complex field.

\section{HHL Algorithm and Swap Test}
\label{HHL Algorithm and Swap Test}

In this section, we will introduce some powerful quantum techniques comprehensively that will be used in this work. It can also be regarded as a brief review of HHL algorithm and swap test. For some basic definitions about quantum computing, we refer to \cite{nielsen}.

\subsection{Quantum phase estimation algorithm}

Quantum phase estimation algorithm is one of the most important techniques in quantum algorithm designing. It was first proposed by Kitaev at 1995 \cite{kitaev} as an extension of Shor's algorithm \cite{shor}. Most important quantum algorithms, such as Shor's factoring and discrete logarithm algorithm \cite{shor}, HHL algorithm to linear system \cite{harrow}, quantum counting \cite{brassard} are based on it. The problem considered in quantum phase estimation algorithm can be stated as:
Let $U$ be a unitary transformation with a given eigenvector $|u\rangle$, then how to find the corresponding eigenvalue $e^{2\pi \i \theta}$ ($0\leq \theta <1$), such that $U|u\rangle=e^{2\pi \i\theta}|u\rangle$.

The designing of this algorithm is not so difficult, which is a beautiful application of quantum Fourier transformation. In the following, integer $n$ is related to the bit precision we want to obtain about $\theta$.
More precisely, the algorithm will find an $\tilde{\theta}$ such that $|\theta-\tilde{\theta}|\leq 2^{-n}$. It also refers to the number of qubit required in this algorithm. Denote $N=2^n$ for simplicity.
The quantum phase estimation algorithm contains four steps:

\emph{Step 1.} Prepare the initial state $|\psi_0\rangle = |0\rangle^{\otimes n} |u\rangle$.

\emph{Step 2.} Apply Hadamard transformation $H^{\otimes n}$ on the first register of $|\psi_0\rangle$ to generate a superposition:
\[
|\psi_1\rangle = \frac{1}{\sqrt{N}}\sum_{x=0}^{N-1}|x\rangle|u\rangle.
\]

\emph{Step 3.} Apply control transformation $\sum_{x=0}^{N-1}|x\rangle\langle x|\otimes U^x$ on $|\psi_1\rangle$, that is applying $U^x$ on $|u\rangle$ if the first register is $|x\rangle$. Then we have
\[
|\psi_2\rangle = \frac{1}{\sqrt{N}}\sum_{x=0}^{N-1}|x\rangle U^x|u\rangle=\frac{1}{\sqrt{N}}\sum_{x=0}^{N-1}e^{2\pi \i\theta x} |x\rangle|u\rangle
\]

\emph{Step 4.} Apply the inverse quantum Fourier transformation on the first register
\[\ba{lll} \vspace{.2cm}
|\psi_3\rangle &=& \ds\frac{1}{N}\sum_{y=0}^{N-1}\sum_{x=0}^{N-1}e^{2\pi \i\theta x-2\pi \i \frac{xy}{N}}|y\rangle |u\rangle \\
               &=& \ds\frac{1}{N}\sum_{y=0}^{N-1}\Bigg[\sum_{x=0}^{N-1}e^{2\pi \i x(\theta-\frac{y}{N})}\Bigg]|y\rangle |u\rangle.
\ea\]

Finally, perform measurements. When $n$ is chosen, we will get a $2^{-n}$ approximtae of $\theta$ with a high probability close to 1. More detailed analysis are given below.
For convenience, denote $\delta(y)=\theta-\frac{y}{N}$. Note that $0\leq\theta<1$, so $\theta$ can be written in binary form as
\be \label{quantum phase estimation:eq1}
\theta=\theta_1\frac{1}{2}+\cdots+\theta_n\frac{1}{2^n}+r_n=\frac{\theta_12^{n-1}+\cdots+\theta_n}{2^n}+r_n,
\ee
where $\theta_i\in\{0,1\}$ and $0\leq r_n\leq 2^{-n}$. For simplicity, we denote $y_\theta=\theta_12^{n-1}+\cdots+\theta_n$.

\emph{Case 1}. If $r_n=0$, then $\delta(y_\theta)=0$ and $|\psi_3\rangle=|y_\theta\rangle|u\rangle$. At this time, we can obtain $y_\theta$ with probability 1 by measurement, and the algorithm is deterministic.

\emph{Case 2}. If $0<r_n<2^{-(n+1)}$, then $\delta(y_\theta)=r_n\leq 2^{-(n+1)}$, and the probability of $|y_\theta\rangle$ is
\be\ba{lll} \label{quantum phase estimation:eq2}  \vspace{.2cm}
\textmd{Prob}(|y_\theta\rangle)
&=& \ds\frac{1}{N^2}\Bigg|\sum_{x=0}^{N-1}e^{2\pi \i x\delta(y_\theta)}\Bigg|^2 \\ \vspace{.2cm}
&=& \ds \frac{1}{N^2}\Bigg|\frac{e^{2\pi \i N\delta(y_\theta)}-1}{e^{2\pi \i \delta(y_\theta)}-1}\Bigg|^2 \\ \vspace{.2cm}
&=& \ds \frac{1}{N^2}\Bigg|\frac{\sin (N\delta(y_\theta)\pi)}{\sin (\delta(y_\theta)\pi)}\Bigg|^2  \\
&\geq& \ds \frac{4}{N^2\pi^2}\frac{N^2\delta(y_\theta)^2\pi^2}{\delta(y_\theta)^2\pi^2}=\frac{4}{\pi^2}.
\ea\ee
Here we use the fact that if $|x|\leq \pi/2$, then $2x/\pi\leq |\sin x|\leq x$. Note that, at this time, the probability of $|y_\theta+1\rangle$ may be small, but $(y_\theta+1)/2^n$ also provides a good approximate of $\theta$ due to $\frac{y_\theta+1}{2^n}-\theta=2^{-n}-r_n\leq 2^{-n}$.

\emph{Case 3}. If $2^{-(n+1)}\leq r_n\leq 2^{-n}$, then at this time $|\delta(y_\theta+1)|=2^{-n}-r_n\leq 2^{-(n+1)}$. Similar to the analysis in \eqref{quantum phase estimation:eq2},  the probability of $|y_\theta+1\rangle$ satisfies
\be
\textmd{Prob}(|y_\theta+1\rangle)\geq\frac{4}{\pi^2}.
\ee
Also, at this time, $y_\theta/2^n$ is a good approximate of $\theta$ even though the probability of $|y_\theta\rangle$ may be small.

Combining the above analysis, we conclude that we have a high probability larger than $4/\pi^2$ to get a $\tilde{y}\in\{y_\theta,y_\theta+1\}$,
such that $|\theta-\frac{\tilde{y}}{2^n}| \leq 2^{-n}$. So we can get two good approximates of $\theta$.

From (\ref{quantum phase estimation:eq1}), we see that $\theta\in[y_\theta/2^n,(y_\theta+1)/2^n]$.
And the above analysis depends on whether $\theta$ is closer to $y_\theta/2^n$ or to $(y_\theta+1)/2^n$.
It bisects the interval $\theta\in[y_\theta/2^n,(y_\theta+1)/2^n]$ into two subintervals.
Note that in (\ref{quantum phase estimation:eq1}), $n$ is the bit accuracy we want to obtain, which means the first $(n-1)$ bits of $\theta$ is determined by $\tilde{y}$ with no error.
Generally, we can approximate $\theta$ to precision $2^{-m}$ instead of $2^{-n}$, i.e., find a $y$ such that $|\frac{y}{2^n}-\theta|\leq2^{-m}$, here $m\leq n$.
We change the expression of (\ref{quantum phase estimation:eq1}) into
\be \label{quantum phase estimation:eq3}
\theta=\frac{\theta_12^{m-1}+\cdots+\theta_m}{2^m}+r_m=:\frac{\alpha_m}{2^m}+r_m.
\ee
Suppose $n=m+p$. Then a similar idea is splitting the interval $[\alpha_m/2^m,(\alpha_m+1)/2^m]$ into $2^{p+1}$ equal parts. And we will get $2^p+1$ good approximates of $\theta$,
i.e.,
\be
\{(2^p\alpha_m+t)/2^n\mid t=0,1,\ldots,2^p\}.
\ee
It has been proved in \cite{nielsen} that the success probability of obtaining these good approximates of $\theta$ is at least $1-1/2(2^p-2)$.
So based on the bit accuracy and the successful probability we want, we can determine the value of $n$. If we denote the precision
$\epsilon=2^{-m}$ and the failure probability $\delta=1/2(2^p-2)$,
then
\be \label{quantum phase estimation:eq4}
n=\lceil\log1/\epsilon\rceil+\lceil\log(2+\frac{1}{2\delta})\rceil=O(\log1/\epsilon\delta).
\ee
Concluding the above analysis, we have

\bp
\cite{kitaev}
Let $U$ be a unitary transformation with implementation complexity $O(T_U)$ and $|u\rangle$ an eigenvector of $U$. Then quantum phase estimation algorithm can find the corresponding eigenvalue in time $O(T_U/\epsilon\delta)$ to precision $\epsilon$ with a successful probability at least $1-\delta$.
\ep

If we only need the successful probability larger than 2/3, then the complexity of quantum phase estimation algorithm just equals $O(T_U/\epsilon)$.

An important advantage of the quantum phase estimation is that we can estimate all eigenvalues $\{e^{2\pi\theta_j}\mid j=1,\ldots,M\}$ of $U$
even without knowing the eigenvectors $\{|u_j\rangle\mid j=1,\ldots,M\}$, where $M$ is the size of $U$.
The idea is pretty similar to above.
Arbitrary choose an initial state $|c\rangle$. We can formally rewrite it as $|c\rangle=\sum_{j=1}^M\gamma_j|u_j\rangle$ due to $\{|u_j\rangle\mid j=1,\ldots,M\}$ forms an orthogonal basis,
where $\gamma_j=\langle c|u_j\rangle$. The procedure is exactly the same as step 1-4, except the initial state becomes $|0\rangle^{\otimes n}|c\rangle$.
The final result is an approximate of
\be \label{quantum phase estimation:eq5}
\sum_{j=1}^M \gamma_j|\theta_j\rangle |u_j\rangle,
\ee
where the first register stores the eigenvalue information and the second register stores the eigenvector information of $U$. The complexity also equals $O(T_U/\epsilon)$. The expression \eqref{quantum phase estimation:eq5} may not rigorous superficially, since there will be some garbage states we do not want in the final state. However, on one hand, based on the analysis about the success probability, the amplitude of \eqref{quantum phase estimation:eq5} in the final state is almost close to 1. This makes the expression  \eqref{quantum phase estimation:eq5} more reasonable.
On the other hand, the expression \eqref{quantum phase estimation:eq5}  contains a perfect performance on intuition about the eigenvalue and eigenvector information about $U$. This will bring a lot of convenience for further use.

\subsection{HHL algorithm}

Combining quantum phase estimation and Hamiltonian simulation,  we can actually estimate the eigenvalues and eigenvectors of Hermitian matrix \cite{abrams}. This forms one central step of HHL algorithm.

Consider the linear system $Ax=b$. We assume that $A$ is a $M\times M$ Hermitian matrix, otherwise we can consider an equivalent linear system
$\left[
   \begin{array}{cc}
     0 & A \\
     A^\dag & 0 \\
   \end{array}
 \right]\left[
          \begin{array}{c}
            0 \\
            x \\
          \end{array}
        \right]=\left[
          \begin{array}{c}
            b \\
            0 \\
          \end{array}
        \right]$.
Since $A$ is Hermitian matrix, $U=e^{\i A t}$ is unitary, which can be efficiently simulated in quantum computer when $A$ is sparse \cite{berry, berry1}.
Suppose $A=\sum_{j=1}^M \sigma_j|u_j\rangle\langle u_j|$ is an eigenvalue decomposition of $A$, then we can formally rewrite $|b\rangle=\sum_{j=1}^M\gamma_j|u_j\rangle$ for some unknown coefficients $\gamma_j$.
We also assume that $1/\kappa\leq |\sigma_i| < 1$, where $\kappa$ is the condition number of $A$, otherwise we can perform a suitable scaling on the original linear system.
By (\ref{quantum phase estimation:eq5}), we can get an approximate of
\be \label{quantum phase estimation:eq6}
\sum_{j=1}^M \gamma_j|\sigma_j\rangle |u_j\rangle,
\ee
in time $O((\log n)/\epsilon)$, since $U$ is efficiently simulated now. Then perform a controlled rotation based on the $\sigma_j$, which yields
\be \label{quantum phase estimation:eq7}
\sum_{j=1}^M \gamma_j|\sigma_j\rangle |u_j\rangle \Bigg[\sigma_j^{-1}|0\rangle+\sqrt{1-\sigma_j^{-2}}|1\rangle\Bigg].
\ee
Finally, undo the quantum phase estimation algorithm,
\be \label{quantum phase estimation:eq8}
\sum_{j=1}^M \gamma_j|u_j\rangle \Bigg[\sigma_j^{-1}|0\rangle+\sqrt{1-\sigma_j^{-2}}|1\rangle\Bigg].
\ee
The first part $\sum_{j=1}^M \gamma_j\sigma_j^{-1} |u_j\rangle$ equals $A^{-1}|b\rangle$. Perform measurements, we will get the quantum state of the solution in time  $O(\kappa^2(\log n)/\epsilon)$. The above is the main idea of HHL algorithm. More detailed analysis is given below:

(a). Quantum phase estimation algorithm only returns an $\epsilon$ approximate say $\tilde{\sigma}_j$ of $\sigma_j$, that is $|\sigma_j-\tilde{\sigma}_j|\leq \epsilon$.
So
\[
\frac{|\sigma_j^{-1}-\tilde{\sigma}_j^{-1}|}{|\sigma_j^{-1}|}
=\frac{|\sigma_j-\tilde{\sigma}_j|}{|\tilde{\sigma}_j^{-1}|} \leq \epsilon\kappa.
\]
Estimating the inverse of eigenvalues will enlarge the error by a factor $\kappa$. This will lead to a factor of condition number in the complexity.

(b). The success probability in (\ref{quantum phase estimation:eq8}) is $\sum_{j=1}^M |\gamma_j\sigma_j^{-1}|^2 \geq \sum_{j=1}^M |\gamma_j|^2/\kappa^2=1/\kappa^2$. Due to amplitude amplification, after $O(\kappa)$ times of measurements, we will have a high probability close to 1 to get the quantum state of the solution.  This will lead to another factor of condition number in the complexity.

(c). The reason why undoing quantum phase estimation does not affect $\sigma_j^{-1}|0\rangle+\sqrt{1-\sigma_j^{-2}}|1\rangle$ is that, quantum phase estimation returns good approximates of the eigenvalues of $A$, so to some sense $\sigma_j$ only depends on $|u_j\rangle$ at step (\ref{quantum phase estimation:eq7}). Quantum phase estimation does not change $|u_j\rangle$, so it will not change the state $\sigma_j^{-1}|0\rangle+\sqrt{1-\sigma_j^{-2}}|1\rangle$ when undoing it.

The following are some further remarks about HHL algorithm:

(a). HHL only returns one solution, i.e., the least square solution. More precisely, the solution of HHL algorithm has the form
$\sum_{j,\sigma_j\neq 0} \gamma_j\sigma_j^{-1} |u_j\rangle$.

(b). HHL algorithm needs efficient preparation of the quantum state $|b\rangle$ of $b$, which can achieved when $b$ is sparse or relatively uniform distributed.

(c). The solution of HHL algorithm is a quantum state $|x\rangle$ of the solution $x$, not a classical solution. Reading out the classical solution takes at least $O(M)$ steps, which kills the exponential speedup of HHL algorithm.
About quantum state $|x\rangle$, currently we can only perform swap test to estimate the inner product of $|x\rangle$ with some other quantum states $|y\rangle$. However, this is already enough to solve many problems.

(d). HHL algorithm requires $A$ to be invertible or $b$ lies in the well-conditioned parts of $A$. Simply, if $\sigma_j=0$, then we should have $\gamma_j=0$. Or the components with $\gamma_j\neq0$ but $\sigma_j=0$ only occupy a small part in $b$.

(e). As a generalization of HHL algorithm, we not only can compute the inverse of $A$, but also can compute any polynomial of $A$ from formula (\ref{quantum phase estimation:eq6}). This is achieved by changing $\sigma_j^{-1}$ into any polynomial of $\sigma_j$.
So matrix multiplication, matrix power and many other matrix operations can achieved efficiently in quantum computer by HHL algorithm. This will help us solve lots of problems relate to matrix.

(f). HHL algorithm needs efficient simulation of Hamiltonian $e^{-\i At}$. This is already solved in the sparse case. Also it is efficient when $A$ is dense but low rank \cite{rebentros16}.

\subsection{Swap test}

Currently, one efficient operation among quantum states is swap test \cite{buhrman}. For any two quantum states $|x\rangle,|y\rangle$, swap test can be used to estimate
${\rm Re}\langle x|y\rangle$ efficiently. By considering $|x\rangle,\i|y\rangle$, we can also get ${\rm Im}\langle x|y\rangle$ efficiently. Estimating inner product is already enough to solve many problems, so this subsection devotes to give a brief analysis about swap test. The following lemma is a result of quantum phase estimation algorithm.

\bl
\label{lem:inner product}
Let $|\phi\rangle=\sin\theta|0\rangle|u\rangle+\cos\theta|1\rangle|v\rangle$ be a unknown quantum state that can be prepared in time $O(T_{\emph{in}})$, where $|u\rangle,|v\rangle$ are normalized quantum states. Then there is a quantum algorithm that can compute $\sin\theta,\cos\theta$ in time $O(T_{\emph{in}}/\epsilon\delta)$ in precision $\epsilon$ with success probability at least $1-\delta$.
\el

{\em Proof.}
Let $Y$ be the 2-dimensional unitary transformation that maps $|0\rangle$ to $-|0\rangle$ and $|1\rangle$ to $|1\rangle$. Denote $G=(2|\phi\rangle\langle\phi|-I)(Y\otimes I)$ which is the rotation matrix used in Grover's searching algorithm. Then $G$ has the following matrix representation
\[G=\left[
   \begin{array}{rr} \vspace{.2cm}
     \cos2\theta  &~~ \sin2\theta \\
     -\sin2\theta &~~ \cos2\theta \\
   \end{array}
 \right]\]
in the space ${\rm span}\{|0\rangle|u\rangle,|1\rangle|v\rangle\}$.
The eigenvalues of $G$ are $e^{\pm \i2\theta}$
and the corresponding eigenvectors are
\[
|w_\pm\rangle=\frac{1}{\sqrt{2}}\Big(|0\rangle|u\rangle\pm\i|1\rangle|v\rangle\Big).
\]
Note that
$
|\phi\rangle = -\frac{\i}{\sqrt{2}}\Big(e^{\i\theta}|w_+\rangle-e^{-\i\theta}|w_-\rangle\Big).
$
So performing quantum phase estimation algorithm on $G$ with initial state $|0\rangle^n|\phi\rangle$, for some $n=O(\log1/\delta\epsilon)$, can help us find an approximation $\tilde{\theta}$ of $\theta$ with failure probability $\delta$, such that $|\tilde{\theta}-\theta|\leq \epsilon$. 
\hfill $\square$

Generally, the failure probability $\delta$ can be ignored. Simply speaking, the above lemma can be used to estimate the amplitude (or probability) of certain states efficiently.
A directly corollary of lemma \ref{lem:inner product} is swap test, which is described as below

\bp
\label{cor:inner product}
Let $|x\rangle,|y\rangle$ be two quantum states, which can be prepared in time $O(T_{\emph{in}})$,
then ${\rm Re}\langle x|y\rangle$ can be estimated in precision $\epsilon$ in time $O(T_{\emph{in}}/\epsilon)$.
\ep

{\em Proof.}
Consider the following procedure:
\[\ba{lll}\vspace{.2cm}
\ds \frac{1}{\sqrt{2}} (|0\rangle+|1\rangle)|0\rangle
&\mapsto& \ds\frac{1}{\sqrt{2}} (|0\rangle|x\rangle+|1\rangle|y\rangle) \\
&\mapsto& \ds\frac{1}{2} |0\rangle(|x\rangle+|y\rangle)+\frac{1}{2} |1\rangle(|x\rangle-|y\rangle).
\ea\]
The first and third step are the result of Hadamard operation on the first qubit. Denote the final quantum state as $|\phi\rangle$.
Then the probability of $|0\rangle$ (resp. $|1\rangle$) equals $(1+{\rm Re}\langle x|y\rangle)/2$ (resp. $(1-{\rm Re}\langle x|y\rangle)/2$).
By lemma \ref{lem:inner product}, these two values can be evaluated in time $O(T_{\textmd{in}}/\epsilon)$ to precision
$\epsilon$.
Then so is ${\rm Re}\langle x|y\rangle$.
\hfill $\square$

As discussed in the beginning, we actually have

\bp  \label{cor:swap test}
For any two quantum states $|x\rangle,|y\rangle$, which can be prepared in time $O(T_{\emph{in}})$, then there is a quantum algorithm to estimate $\langle x|y\rangle$ to precision $\epsilon$ in time $O(T_{\emph{in}}/\epsilon)$.
\ep

\section{Quantum State Preparation}
\label{Quantum State Preparation}

Let $x=(x_0,\ldots,x_{m-1})$ be a complex vector, the quantum state it corresponds to equals $|x\rangle=\frac{1}{\|x\|} \sum_{i=0}^{m-1} x_i|i\rangle$. The transformation from classical data $x$ into its quantum state $|x\rangle$ is usually called the ``input problem" in quantum computer \cite{biamonte}, which forms the initial step in many quantum algorithms, such as
\cite{childs-linear-system,clader,harrow,kerenidis,kerenidis-iteration,lloyd13,rebentros17-Network, rebentros14,rebentros17b, rebentros16, wang,wiebe,wossnig}. It is also important in this work. In the following, we show one method to do this job based on linear combination of uniatries (LCU for short), which achieves an exponential speedup than the algorithm given in \cite{clader}.

The LCU problem can be stated as: given $m$ complex numbers $\alpha_j$ and $m$ quantum states $|x_j\rangle$, which can be prepared efficiently in time $O(T_{\textmd{in}})$, where $j=0,1,\ldots,m-1$, then how to prepare the quantum state $|y\rangle$ proportional to $y=\sum_{j=0}^{m-1} \alpha_j |x_j\rangle$? And what is the corresponding efficiency? LCU was first proposed by Long \cite{long06, long11}.
In the following, we focus on one simple form \cite{childs-linear-system}.

Set $\alpha_j=r_je^{\i\theta_j}$, where $r_j>0$ is the norm of $\alpha_j$. Denote $s=\sum_{j=0}^{m-1}r_j$. Define unitary transformation $S$ as $S|0\rangle=\frac{1}{\sqrt{s}}\sum_{j=0}^{m-1}\sqrt{r_j}|j\rangle$.
Then $|y\rangle$ can be obtained from the following procedure:
\be\ba{lcl}\vspace{.2cm} \label{lcu2}
|0\rangle|0\rangle &\xrightarrow[]{S\otimes I}& \ds\frac{1}{\sqrt{s}} \sum_{j=0}^{m-1} \sqrt{r_j}|j\rangle|0\rangle \\\vspace{.2cm}
&\rightarrow& \ds\frac{1}{\sqrt{s}} \sum_{j=0}^{m-1} \sqrt{r_j}e^{\i\theta_j}|j\rangle|x_j\rangle \\
&\xrightarrow[]{S^\dagger\otimes I}& \ds\frac{1}{s} |0\rangle \sum_{j=0}^{m-1} \alpha_j|x_j\rangle +\textmd{orthogonal parts}.
\ea\ee
The second step is a control operation to prepare $|x_j\rangle$ with respect to $|j\rangle$.
The probability to get $|y\rangle$ equals $\|y\|^2/s^2$, and so the complexity to obtain $|y\rangle$  is
$O((T_{\textmd{in}}+\log m)s/\|y\|)$. A direct corollary of this LCU is

\bp\label{state-preparation1}
For any vector $x=(x_0,\ldots,x_{m-1})$, its quantum state can be prepared in time $O(\kappa(x)\log m)$, where $\kappa(x)=\max_k|x_k|/\min_{k,x_k\neq 0}|x_k|$.
\ep

{\em Proof.}
We assume that all entries of $x$ are nonzero, otherwise we only focus on the nonzero entries of $x$.
Then it suffices to choose $|x_j\rangle=|j\rangle$ in \eqref{lcu2}. At this time $O(T_{\rm in})=O(1)$. So the complexity is
$O((T_{\textmd{in}}+\log m)\sum_j|\alpha_j|/\|y\|)=O(\kappa(x)\log m)$, since $\|y\|\geq m\min_j|x_j|$ and $s\leq m\max_j|x_j|$.
\hfill $\square$

Actually, based on LCU, the quantum state can be prepared more efficiently.

\bt \label{state-preparation2}
Let $x=(x_0,\ldots,x_{m-1})$ be a given vector and $\kappa(x)=\max_k |x_k|/\min_{k,x_k\neq 0} |x_k|$. Then the quantum state of $x$ can be prepared in time $O(\sqrt{\log\kappa(x)}\log m)$.
\et

{\em proof}
For simplicity, we assume that $|x_0|=\min_{k,x_k\neq 0}|x_k|$. Find the minimal $q$ such that $\kappa(x)\leq 2^q$, so $q\approx \log \kappa(x)$. For any $1\leq j\leq q$, there are several entries of $x$ such that their absolute values lie in the  interval $[2^{j-1}|x_0|,2^{j}|x_0|)$. Define $y_j$ as the $n$ dimensional vector by filling these entries into the corresponding positions as them in $x$ and zero into other positions. Then $x=y_1+\cdots+y_q$. For any $j$, we have $\kappa(y_j)\leq 2$, so the quantum state $|y_j\rangle$ of vector $y_j$ can be prepared efficiently in time $O(\log m)$ by proposition \ref{state-preparation1}. We also have
$|x\rangle=\lambda_1|y_1\rangle+\cdots+\lambda_q|y_q\rangle$, where $\lambda_j=\|y_j\|/\|x\|$. From the LCU method (\ref{lcu2}) given above, the complexity to achieve such a linear combination to get $|x\rangle$ equals
$
O(\log m\sum_{j=1}^q {\|y_j\|}/{\|x\|})
= O(\sqrt{q}\log m) = O(\sqrt{\log\kappa(x)}\log m) ,
$
where the first identity is because of the relation between 1-norm and 2-norm of vectors, more precisely, it is a result of
$\sum_{j=1}^q \|y_j\| \leq \sqrt{q} \sqrt{\sum_{j=1}^q \|y_j\|^2}=\sqrt{q}\|x\|$.
\hfill $\square$

The quantum algorithm to prepare quantum states given in \cite{clader} is based on another LCU method, which can be viewed as an inspiration of HHL algorithm. The corresponding complexity is the same as proposition \ref{state-preparation1}. Note that the quantum algorithm used to study supervised classification \cite{lloyd13} also induces a method to prepare quantum states, the complexity is a little worse than proposition \ref{state-preparation1}. Compared with these two works, the new quantum algorithm actually achieves an exponential speedup in $\kappa(x)$.

\section{Preliminaries of Cubic Spline Interpolation}
\label{Preliminaries of Cubic Spline Interpolation}

In this section, we briefly review the cubic spline interpolation method, more details can be found in \cite{burden, li, sauer}. Since the aim of this work is providing a new application of HHL algorithm, we will not go deeper about cubic spline interpolation and its applications or generalizations. Given a data set of $n+1$ samples
\[\mathcal{X}=\{(x_i,y_i):i=0,1,\ldots,n~\textmd{and}~x_i\neq x_j~\textmd{if}~i\neq j\},\]
where $x_i,y_i\in \mathbb{R}$. We also assume that $a=x_0<x_1<\cdots<x_n=b$. The spline function $S(x)$ is a function satisfying:
\begin{enumerate}
  \item $S(x)$ is second differentiable in the interval $[a,b]$;
  \item $S(x)$ is a polynomial of degree 3 in each subinterval $[x_i,x_{i+1}]$ for all $i=0,1,\ldots,n-1$;
  \item $S(x_i)=y_i$ for all $i=0,1,\ldots,n$.
\end{enumerate}

Because of condition 2, we denote the cubic polynomial in subinterval $[x_i,x_{i+1}]$ as $C_i(x)$. Then there are totaly $4n$ unknown parameters we should determine in $S(x)$. By condition 1 and 3, we have the following $4n-2$ conditions:
\be \label{eq0}
\left\{
  \begin{array}{ll} \vspace{.2cm}
    C_i(x_i)=y_i~\textmd{and}~C_i(x_{i+1})=y_{i+1}, & \hbox{} \\ \vspace{.2cm}
    C_i'(x_{i+1})=C_{i+1}'(x_{i+1}), & \hbox{} \\
    C_i''(x_{i+1})=C_{i+1}''(x_i). & \hbox{}
  \end{array}
\right.
\ee
Usually we will add two extra boundary conditions to make the spline function unique. There are three types of frequently used boundary conditions:

{\bf Type 1.} The first derivatives of $S(x)$ at the endpoints are known:
  \be \label{type1}
  C_0'(x_0)=f_0'~\textmd{and}~C_{n-1}'(x_n)=f_n'.
  \ee
  The special case $C_0'(x_0)=C_{n-1}'(x_n)=0$ will be called \emph{clamped boundary conditions}.

{\bf Type 2.} The second derivatives of $S(x)$ at the endpoints are known:
  \be \label{type2}
  C_0''(x_0)=f_0''~\textmd{and}~C_{n-1}''(x_n)=f_n''.
  \ee
  The special case $C_0''(x_0)=C_{n-1}''(x_n)=0$ will be called \emph{natural boundary conditions}.

{\bf Type 3.} Since cubic spline interpolation can be used to approximate a given function $f(x)$. At this case, the input data $\mathcal{X}$ are given in the form $y_i=f(x_i)$. When the exact function $f(x)$ is a periodic function with period $x_n-x_0$, we also need $S(x)$ to be a periodic function with period $x_n-x_0$. Thus the required conditions include
  \be\ba{lll} \vspace{.2cm} \label{type3}
  \left\{
    \begin{array}{ll} \vspace{.2cm}
      C_0(x_0)=C_{n-1}(x_n), & \hbox{} \\ \vspace{.2cm}
      C_0'(x_0)=C_{n-1}'(x_n), & \hbox{} \\
      C_0''(x_0)=C_{n-1}''(x_n). & \hbox{}
    \end{array}
  \right.
  \ea\ee
  The spline function $S(x)$ in this type is called \emph{periodic splines}.

There are several typical methods that can be used to find the spline function $S(x)$ according to its corresponding conditions \cite{burden, li, sauer}. The main ideas are the same. In the following, we follow the idea of \cite{li} by considering the second derivatives $S''(x_i)=M_i~(i=0,1,\ldots,n)$ as the initial step. The problem now reduces to compute all $M_i$. By Lagrange interpolation with the boundary condition $C_i''(x_i)=M_i$ and $C_i''(x_{i+1})=M_{i+1}$, we can interpolate each $C_i''$ on interval $[x_i,x_{i+1}]$ in the following form
\be \label{eq1}
C_i''(x)=M_i\frac{x_{i+1}-x}{h_i}+M_{i+1}\frac{x-x_i}{h_i},
\ee
where
$
h_i=x_{i+1}-x_i.
$
Integrating the equation (\ref{eq1}) twice and using the conditions $C_i(x_i)=y_i$ and $C_i(x_{i+1})=y_{i+1}$, we have
\be\ba{lll} \vspace{.2cm} \label{csi:eq1}
C_i(x) &=& \ds \frac{M_i}{6h_i}(x_{i+1}-x)^3+\frac{M_{i+1}}{6h_i}(x-x_i)^3 \\ \vspace{.2cm}
    & & +\ds \left(y_i-\frac{M_ih_i^2}{6}\right)\frac{x_{i+1}-x}{h_i} \\
    & & +\ds \left(y_{i+1}-\frac{M_{i+1}h_i^2}{6}\right)\frac{x-x_i}{h_i}.
\ea\ee
Therefore,
\be\ba{rll} \vspace{.2cm}
C'_i(x_{i+1})     & =& \ds \frac{(M_i+2M_{i+1})h_i}{6}+\frac{y_{i+1}-y_i}{h_i}, \\
C'_{i+1}(x_{i+1}) & =& \ds-\frac{(2M_{i+1}+M_{i+2})h_{i+1}}{6}+\frac{y_{i+2}-y_{i+1}}{h_{i+1}}.
\ea\ee
These two values should equal to each other because of the second equality in formula (\ref{eq0}), so
\be \label{eq3}
\mu_{i+1} M_i+2M_{i+1}+\lambda_{i+1} M_{i+2} = d_{i+1},
\ee
where for any $i=0,1,\ldots,n-2,$
\be\ba{lll} \vspace{.2cm} \label{parameter0}
&& \ds \mu_{i+1}=\frac{h_i}{h_i+h_{i+1}}, \\\vspace{.2cm}
&& \ds \lambda_{i+1}=1-\mu_{i+1}=\frac{h_{i+1}}{h_i+h_{i+1}},  \\
&& \ds d_{i+1}=6S[x_i,x_{i+1},x_{i+2}].
\ea\ee
Here $S[x_i,x_{i+1},x_{i+2}]$ is the \emph{Newton divided difference}. It is defined recursively,
\[\ba{rll}\vspace{.2cm}
\ds S[x_i,x_{i+1},x_{i+2}] &=& \ds\frac{S[x_{i+1},x_{i+2}]-S[x_i,x_{i+2}]}{x_{i+2}-x_i}, \\
\ds S[x_i,x_{i+1}] &=& \ds\frac{S(x_{i+1})-S(x_i)}{x_{i+1}-x_i},
\ea\]
with initial values $S(x_i)=y_i$.

For type 1 boundary condition, we will have
\be\ba{rll} \vspace{.2cm}
2M_0+M_1     &=& \ds \frac{6}{h_0}(S[x_0,x_1]-f_0'), \\
M_{n-1}+2M_n &=& \ds \frac{6}{h_{n-1}}(f_n'-S[x_{n-1},x_n]).
\ea\ee
Hence,
we can set $\lambda_0=\mu_n=1$,
$d_0=6(S[x_0,x_1]-f_0')/h_0$ and
$d_n=6(f_n'-S[x_{n-1},x_n])/h_{n-1}$.
Finally, the linear system of equations that we need to solve has the form
\be \label{matrix1}
\left[{\begin{array}{ccccccc}
2       &\lambda_0 &          &&\\
\mu_1   &2         &\lambda_1 &&\\
        &\ddots    &\ddots    &\ddots & \\
&       &\mu_{n-1} &2         &\lambda_{n-1}\\
&       &          &\mu_{n}   &2
\end{array}}\right]
\left[
  \begin{array}{c}
    M_0 \\
    M_1 \\
    \vdots \\
    M_{n-1} \\
    M_n \\
  \end{array}
\right]=\left[
  \begin{array}{c}
    d_0 \\
    d_1 \\
    \vdots \\
    d_{n-1} \\
    d_n \\
  \end{array}
\right].
\ee

For type 2 boundary condition, we have $M_0=f''_0$ and $M_n=f''_n$, so we can set $\lambda_0=\mu_n=0$ and $d_0=2f''_0, d_n=2f''_n$.
Then we need to solve a linear system in the same form as above with different values at $\lambda_0,\mu_n,d_0,d_n$.

For type 3 boundary condition, we have
\be
M_0=M_n,~~\lambda_n M_1+\mu_n M_{n-1}+2M_n=d_n,
\ee
where
\be\ba{lll}\vspace{.2cm} \label{paramters1}
&& \ds \lambda_n=\frac{h_0}{h_{n-1}+h_0}, \\\vspace{.2cm}
&& \ds \mu_n=1-\lambda_n=\frac{h_{n-1}}{h_{n-1}+h_0}; \\
&& \ds d_n=6\frac{S[x_0,x_1]-S[x_{n-1},x_n]}{h_0+h_{n-1}}.
\ea\ee
So the linear system we need to solve is
\be \label{matrix2}
\left[{\begin{array}{ccccccc}
2       &\lambda_1 &          &&\mu_1\\
\mu_2   &2         &\lambda_2 &&\\
        &\ddots    &\ddots    &\ddots & \\
&       &\mu_{n-1} &2         &\lambda_{n-1}\\
\lambda_n&         &          &\mu_{n}   &2
\end{array}}\right]
\left[
  \begin{array}{c}
    M_1 \\
    M_2 \\
    \vdots \\
    M_{n-1} \\
    M_n \\
  \end{array}
\right]=\left[
  \begin{array}{c}
    d_1 \\
    d_2 \\
    \vdots \\
    d_{n-1} \\
    d_n \\
  \end{array}
\right].
\ee

The linear system (\ref{matrix1}) is a tridiagonal linear system whose coefficient matrices is diagonally dominant.
The linear system (\ref{matrix2}) is close to a diagonally dominant tridiagonal linear system, except the two values $\mu_1,\lambda_n$.
These two linear systems are very stable and has a unique solution. The classical algorithm, such as Gaussian elimination or the chasing method, to solve such linear systems is not difficult. The complexity is $O(n)$. In this special case, we will believe that quantum computer can achieve exponential speedup by HHL algorithm.

\section{Bounds on Condition Number}
\label{Bounds on Condition Number}

In this section, we focus on the analysis about the condition number of matrices given in (\ref{matrix1}) and (\ref{matrix2}).
The Gershgorin type of circle theorem also holds for singular values \cite{qi}. Let $A=(a_{ij})_{n\times n}$ be any complex matrix, denote
\[r_i=\sum_{\substack{1\leq j\leq n \\ j\neq i}}|a_{ij}|,~~c_j=\sum_{\substack{1\leq i\leq n \\ i\neq j}}|a_{ij}|,~~s_i=\max(r_i,c_i).\]
Then all the singular values of $A$ lie in the following interval
\be \label{sv}
\bigcup_{i=1}^n~[\max(0,|a_{ii}|-s_i),|a_{ii}|+s_i].
\ee

In the linear system (\ref{matrix1}) and (\ref{matrix2}), since $\lambda_i+\mu_i=1$ and $\lambda_i,\mu_i\geq 0$, all the singular values of the coefficient matrices of the linear system (\ref{matrix1}) and (\ref{matrix2}) lie in the following interval by (\ref{sv}),
\be \label{sv:eq1}
\bigcup_{i=1}^n~[2-s_i,2+s_i].
\ee
%In type 1, $\lambda_0=\mu_n=1>\lambda_i,\mu_i>0$ for all $i=1,\ldots,n-1$, so
%\be \label{sv:eq2}
%\max_i(s_i)\leq \max_{1\leq i\leq n-3}\{1+\mu_2,\lambda_i+\mu_{i+2},1+\lambda_{n-2}\}.
%\ee
%In type 2, $\lambda_0=\mu_n=0$ and $1>\lambda_i,\mu_i>0$ for all $i=1,\ldots,n-1$, so
%\be \label{sv:eq3}
%\max_i(s_i)\leq \max_{1\leq i\leq n-3}\{1,\lambda_i+\mu_{i+2}\}.
%\ee
%In type 3, $\lambda_0=\mu_n=0$ and $1>\lambda_i,\mu_i>0$ for all $i=1,\ldots,n-1$, so
%\be \label{sv:eq4}
%\max_i(s_i)\leq \max_{1\leq i\leq n-2}\{1,\lambda_i+\mu_{i+2},\lambda_{n-1}+\mu_1,\lambda_n+\mu_2\}.
%\ee
%The right side of (\ref{sv:eq2}), (\ref{sv:eq3}), (\ref{sv:eq4}) will be denoted as $s$.
In each case, $s_i\leq 2$.
Denote the coefficient matrix of  (\ref{matrix1}) as $A$ and its maximal and minimal singular value as $\sigma_{\max}$ and $\sigma_{\min}$ respectively. Then the above analysis shows that $\sigma_{\max}\leq 4$. Since the coefficient matrix of  (\ref{matrix1}) is invertible, we also have $\sigma_{\max}>0$.
The Frobenius norm of the coefficient matrix of $A$ satisfies:
\[
\|A\|_F^2 = 4n+\sum_{i=0}^{n-1} \lambda_i^2+\sum_{i=1}^{n} \mu_i^2
\geq 4n+\sum_{i=1}^{n-1} (\lambda_i^2+\mu_i^2) \geq \frac{9n}{2}-\frac{1}{2},
\]
where in the last step, we apply the inequality $\lambda_i^2+\mu_i^2\geq 0.5(\lambda_i+\mu_i)^2=0.5$.
By the result about the lower bound of minimal singular value given in \cite{katerina}, we have
\[
\sigma_{\min}^2 \geq  \frac{\|A_n\|_F^2-n\sigma_{\max}^2}{n(1-\sigma_{\max}^2/\det(A)^{2/n})}
\geq \ds \frac{\frac{1}{2}\det(A)^{2/n}}{\det(A)^{2/n}-\sigma_{\max}^2} \geq \frac{1}{2}.
\]
So the condition number of $A$ satisfies $\sigma_{\max}/\sigma_{\min}\leq 4\sqrt{2}$. This result also holds for the case  (\ref{matrix2}). Therefore, the condition number of these two linear systems is bounded by a small constant.
Actually, numerical tests show that, whatever the value of $h_i\geq 0$ is, the condition number of these two linear systems is bounded by 4.

\section{Quantum Cubic Spline Interpolation}
\label{Quantum Cubic Spline Interpolation}

For the linear systems (\ref{matrix1}) and (\ref{matrix2}), the condition number is not too large. Also the coefficient matrix is sparse.
Based on proposition \ref{state-preparation2}, the complexity of the quantum state of the right side of linear systems (\ref{matrix1}) and (\ref{matrix2}) is determined by the value of $\log (\max_k|d_k|/\min_{k,d_k\neq 0}|d_k|)$.
In cubic spline interpolation, the error is controlled by some power of the maximal length of intervals  $\max_k h_k$, which means $h_k$ cannot too large. However, they cannot too small either, otherwise it will bring other troubles in interpolation.
So we can believe that the length $h_k=O(1)$. By definition (\ref{parameter0}), (\ref{paramters1}), the size of $d_k$ is determined by the value of $y_i$ and boundary values (\ref{type1}),  (\ref{type2}), (\ref{type3}). If $n$ is small, then the difference between $\max_k|d_k|$ and $\min_{k,d_k\neq 0}|d_k|$ cannot too large, so the quantum state preparation is efficient. If $n$ is large, then $\max_k|d_k|/\min_{k,d_k\neq 0}|d_k|$ may be very large, however, by taking its logarithm value, the value will be decreased enormously, so we can also believe that the quantum states of the right side hand vectora of linear systems  (\ref{matrix1}) and (\ref{matrix2}) can be prepared efficient. Therefore, all the three restrictions can be solved efficiently in cubic spline interpolation.

The classical algorithm to solve the linear system (\ref{matrix1}) and (\ref{matrix2}) takes time $O(n)$. However, by HHL algorithm, these two linear systems can be solved in time $O((\log n)/\epsilon)$. And we will get a quantum state of the solution
$|M\rangle\varpropto\sum_{i=0}^n M_i|i\rangle.$
Just like linear regression, we can also do further prediction on the new data efficiently. More precisely, suppose we are given a new value $\tilde{x}$. Assume that $\tilde{x}\in[x_i,x_{i+1}]$, then $S(\tilde{x})=C_i(\tilde{x})$. By formula (\ref{csi:eq1}), we have
\be\ba{lll} \vspace{.2cm}
C_i(\tilde{x}) &=& \ds M_i\Bigg[\frac{(x_{i+1}-\tilde{x})^3}{6h_i}-\frac{h_i(x_{i+1}-\tilde{x})}{6}\Bigg] \\ \vspace{.2cm}
& & \ds +M_{i+1}\Bigg[\frac{(\tilde{x}-x_i)^3}{6h_i}-\frac{h_i(\tilde{x}-x_i)}{6} \Bigg] \\ \vspace{.2cm}
& & +\ds y_i\frac{x_{i+1}-\tilde{x}}{h_i}+y_{i+1}\frac{\tilde{x}-x_i}{h_i} \\
&\triangleq& M_i X_i+M_{i+1} X_{i+1}+Y_i.
\ea\ee
Then we just need to prepare the quantum state
$|X\rangle\varpropto X_i|i\rangle+X_{i+1}|i+1\rangle.$
Certainly, this quantum state can be obtained efficiently. By swap test, we can evaluate the inner product of $|M\rangle$ and $|X\rangle$, and so evaluate $S(\tilde{x})$ efficiently in time $O((\log n)/\epsilon^2)$ in precision $\epsilon$. Or on the other hand, we can just apply swap test to evaluate $M_i$, $M_{i+1}$ and $M_{i+2}$, then according to formula (\ref{eq3}) to find out the missed normalization factor. Within the same complexity, we can evaluate $S(\tilde{x})$. Moreover, we can compute the first and second derivatives $S'(\tilde{x})$ and $S''(\tilde{x})$ of $S(x)$ at $\tilde{x}$ within the same time.

%\section{Conclusions}
%
%In this work, we give an efficient quantum algorithm to cubic spline interpolation based on HHL algorithm. It is interesting to find that the obtained quantum algorithm contains no restrictions. There are lots of applications of cubic spline, so the next thing we should focus on is applying this quantum algorithm to solve certain specific important problems that cubic spline interpolation can play a role in.
%
%%\section{Acknowledgements}
%{\bf Acknowledgements.}
%This work is supported by the NSFC Project 11671388 and the CAS Frontier Key Project QYZDJ-SSW-SYS022.

\end{document}